\documentclass[12pt]{iopart}

\usepackage{graphicx}
\usepackage{float}
\usepackage[english]{babel}
\usepackage{amsmath}
\usepackage{dcolumn}
\usepackage{bm}
\usepackage{tikz}
\usepackage{cite}
\usetikzlibrary{decorations.pathreplacing}
\begin{document}

\title[]{Engineering phase and density of Bose-Einstein 
condensates in curved  waveguides with toroidal topology}

\author{Yelyzaveta Nikolaieva$^1$, Luca Salasnich$^{2,3,4}$, Alexander Yakimenko$^{2,3,5}$ \footnote{oleksandr.yakymenko@unipd.it} }
\address{$^1$Vienna Center for Quantum Science and Technology, Atominstitut, TU Wien, Stadionallee 2, 1020 Vienna, Austria }

\address{$^2$Dipartimento di Fisica e Astronomia ’Galileo Galilei’ and Padua QTech Center, Universit{\`a} di Padova, via Marzolo 8, 35131 Padova, Italy}
\address{$^3$Istituto Nazionale di Fisica Nucleare (INFN), Sezione di Padova, via Marzolo 8, 35131 Padova, Italy}
\address{$^4$Istituto Nazionale di Ottica (INO) del Consiglio Nazionale delle Ricerche (CNR), via Nello Carrara 1, 50019 Sesto Fiorentino, Italy}
\address{$^5$Department of Physics, Taras Shevchenko National University of Kyiv,
64/13, Volodymyrska Street, Kyiv 01601, Ukraine}

\vspace{10pt}

\begin{abstract}
   We investigate the effects of ellipticity-induced curvature on atomic Bose-Einstein condensates confined in quasi-one-dimensional closed-loop waveguides. Our theoretical study reveals intriguing phenomena arising from the interplay between curvature and interactions. Density modulations are observed in regions of high curvature, but these modulations are suppressed by strong repulsive interactions. Additionally, we observe phase accumulation in regions with the lowest curvature when the waveguide with superflow has high eccentricity. Furthermore, waveguides hosting vortices exhibit dynamic transformations between states with different angular momenta. These findings provide insights into the behavior of atomic condensates in curved waveguides, with implications for fundamental physics and quantum technologies. The interplay between curvature and interactions offers opportunities for exploring novel quantum phenomena and engineering quantum states in confined geometries.
\end{abstract}

\section{Introduction}
Bose-Einstein condensates (BECs) of atomic gases are fascinating systems that have opened up new perspectives in physics over the past few decades. Investigating atomic BECs has been motivated by researching of fundamental physics of quantum matter and by exploring novel phenomena that arise in the realm of extremely low temperatures at the macroscopic level. Moreover, BECs have practical applications in a variety of fields, including quantum sensing, quantum communication, and quantum computation. These applications are based on the ability to manipulate and control the properties of BECs, such as their coherence, density, and inter-atomic interactions. One of the challenges in BEC research is to manipulate and control the properties of these systems in a precise and robust way. In particular, the effects of external potentials and interactions on the phase and density of BECs are of great interest, as they can induce novel phenomena such as vortices, solitons, and quantum phase transitions \cite{pitaevskii2016bose, pethick2008bose}.

One possible way to manipulate and control BECs is to confine them in curved waveguides, which are quasi-one-dimensional (quasi-1D) and quasi-two-dimensional (quasi-2D) structures (for recent review see \cite{Salasncih_Review2023} and references therein). The influence of curvature on condensate properties has been extensively investigated in both experimental studies of quasi-2D manifolds \cite{2022Natur.606..281C,2023QS&T....8b4003L} and theoretical investigations \cite{Salasnich2019,Salasnich2020,2020NJPh...22f3059M,2020PhRvA.101e3606G,2019npjMG...5...30L, liang2022effective}. Quasi-1D  BECs have attracted significant interest due to their ability to exhibit diverse nonlinear excitations, including dark solitons \cite{2016NJPh...18b5004G,2023JPCM...35a4004M,2018PhRvA..98d3612M,PhysRevA.93.023601,2013NJPh...15k3028B} and solitonic vortices \cite{2015PhRvL.115q0402S,PhysRevA.65.043612,2003PhRvA..68d3617K,2001JPhB...34L.113B,2014PhyOJ...7...82C,2015EPJST.224..577T,2017NJPh...19b3029T}. Curved waveguides can be realized by using magnetic or optical fields to create trapping potentials with different shapes, such as rings, ellipses, or spirals. 
The curvature of these waveguides can modify the phase and density of BECs due to two main effects: the centrifugal force and the geometric potential \cite{olshanii1998atomic}. The centrifugal force arises from the acceleration of atoms along the curved trajectory, while the geometric potential originates from the variation of the transverse confinement along the waveguide. 

In addition to curvature, the geometric potential can be further manipulated by introducing inhomogeneities in the confinement potential along the waveguide. This leads to an effective quantum curvature-induced potential, which exhibits a strong renormalization of the classical curvature-induced potential and significantly enhances the effects of curvature by several orders of magnitude \cite{2001PhRvA..64c3602L,Stringari2006,SciPostPhysCore,Salasncih_Review2023}. The presence of the effective quantum curvature-induced potential gives rise to bound states and energy shifts in curved waveguides \cite{ortix2011curvature}, as well as novel transport phenomena, including coherent backscattering \cite{liang2022effective}.

 Among the simplest and experimentally accessible curved closed-loop quasi-1D waveguides are elliptical waveguides. These waveguides can be created by optical trapping potential or applying a quadrupole magnetic field \cite{gupta2005bose}. Importantly, elliptical waveguides possess a constant eccentricity along their perimeter, enabling the isolation of curvature effects from other factors, such as boundary conditions or nonlinearity.

In waveguides with toroidal topology, the investigation of superfluid flows with nonzero angular momentum holds great significance. These flows are characterized by a nonzero topological charge $q$, which denotes the number of times the phase winds around $2\pi$ along a closed path.
The quantized flow of atomic BECs in closed circuits, have attracted considerable attention due to their relevance in fundamental studies of superfluidity and their potential applications in high precision metrology and atomtronics \cite{Amico2017,Amico2021,Amico2022}. The quantized circulation in a ring corresponds to a $q$-charged vortex line pinned at the center of the ring-shaped condensate, where the vortex energy reaches a local minimum. The confinement provided by the potential barrier surrounding the vortex core makes even multi-charged ($q > 1$) metastable vortex states highly robust.
The generation and stability of these atomic flows in condensates with toroidal topology have been extensively explored both experimentally \cite{Wright2000,Ryu2007,Ramanathan2011,Moulder2012,Ryu2013,Wright2013PhRvL,Wright2013,Murray2013,Beattie2013,Corman2014,Aidelsburger2017,Eckel2014,Eckel2016,Cai2022,DelPace2022} and theoretically \cite{2020PhRvA.102f3324E,Yakimenko2013,Yakimenko2015PhRvA1,Yakimenko2015PhRvA2,2015PhRvA..91f3625M,Snizhko2016,Oliinyk2019JPhB,Oliinyk2019Symm,Oliinyk2020,Yatsuta2020,Bazhan2022,Bland2022}. These investigations have revealed their topological protection in the absence of external driving, highlighting their robust nature.

In this study, we present a comprehensive analysis of steady states in quasi-1D elliptical waveguides with varying eccentricities. We explore the impact of the nonlinear interaction strength on
the density distribution of the stationary ground states, using an approximate 1D model based on the non-polynomial Schr\"odinger equation (NPSE) with an effective curvature-induced potential. Our results are found to be in good qualitative agreement with the numerical solutions of 3D Gross-Pitaevskii equation (GPE). We find that the eccentricity-induced curvature of elliptical waveguides can give rise to two local density peaks in the region with the highest curvature, but repulsive self-interaction counteracts
the resulting curvature-induced density modulation. 
Next, we investigate the effect of curvature on the phase of elliptic waveguides with superflows. We find that high eccentricity results in phase accumulation of the steady states in the regions with the lowest curvature. To study the evolution of superflows with different topological charges, we perform a series of numerical simulations in the framework of damped 3D GPE. Our results demonstrate that elliptical waveguides can provide a versatile tool to manipulate and engineer the properties of BECs in curved geometries. They also reveal new aspects of quantum hydrodynamics in curved manifolds. 

The paper is organized as follows. In section \ref{sec:s0}, we present the results for the stationary ground states in elliptical waveguides. In section \ref{sec:s3}, we present the results for the superflows and their dynamical evolution in elliptical waveguides. In section \ref{sec:Conclusions} we make conclusions.

\section{Stationary ground states of BEC in elliptic waveguide}\label{sec:s0}

\subsection{Ground steady-states of the 3D elliptic BEC}

The Gross-Pitaevskii equation (GPE) is a mean-field approximation that describes the properties of a trapped BEC. In three spatial dimensions, the GPE is given by:
\begin{equation}
i\hbar\frac{\partial \Psi }{\partial t} = \left(-\frac{\hbar^2}{2m} \nabla ^{2}+V_{\text{ext}}(\mathbf{r},t)+g|\Psi |^{2} \right) \Psi. \label{GPE3Ddim}
\end{equation}
Here, $\Psi(\mathbf{r},t)$ represents the wave function of the condensate. The mass of the $^{87}$Rb atom is denoted as $m=1.445\cdot 10^{-25}$ kg, and $\hbar$ represents the reduced Planck's constant. In our investigation, we focus on two scenarios: the non-interacting condensate with $g=0$, and the repulsive interaction case with $g=g_{3D}=4\pi a_{s}\hbar^{2}/m$. For the latter case, the nonlinearity strength $g$ is determined by the $s$-wave scattering length $a_s$, which has a value of $a_{s}=5.31\cdot10^{-9}$ m for $^{87}$Rb atoms. 
To ensure proper normalization, the wave function satisfies the condition $\int|\Psi|^2 d\mathbf{r}=N$, where $N=10^4$ denotes the number of particles in the condensate.
\begin{figure}[ht]
\includegraphics[width=\textwidth]{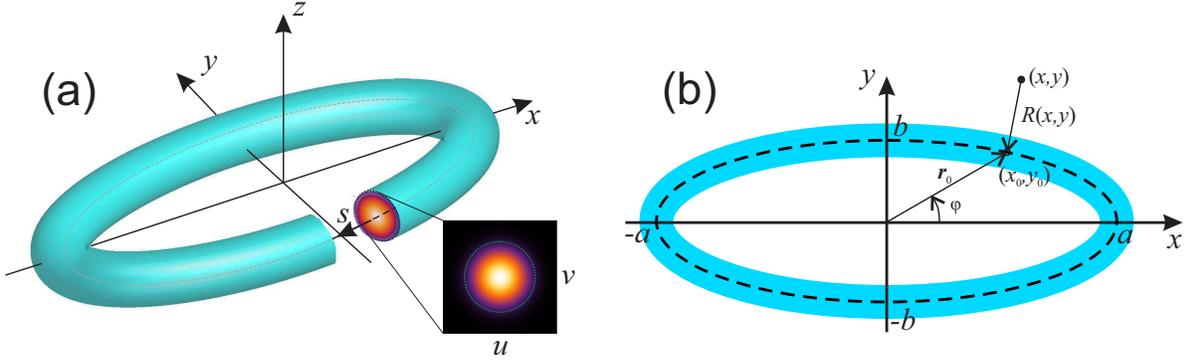}
\caption{Schematics of the elliptic waveguide geometry used for trapping a condensate. (a) 3D plot of the density isosurface and a condensate density in a perpendicular cross-section (brighter colors indicate higher condensate density). (b) The structure of the 3D trapping potential in the $(x, y)$ plane. The minimum points $(x_0, y_0)$ of the potential trap form an ellipse (depicted by a black dashed line). The minimal distance from a point $(x, y)$ to the co-planar point at the ellipse is denoted by $R(x, y)$.}
\label{fig:ellipse_geometry}
\end{figure}

We model the external trapping potential $V_\mathrm{ext}(x,y,z)$ as a combination of an parabolic potential in the $z$-direction and an elliptic waveguide in the $(x,y)$ plane, with a larger semi-axis $a=100\mu$m:
\begin{equation}\label{Eq:V_ext}
V_\mathrm{ext}(x,y,z)=\frac12 m\omega_z^2 z^2+\frac12 m\omega_\perp^2 R^2(x,y).
\end{equation}
Here, $R(x,y)=\left[(x-x_0)^2+(y-y_0)^2\right]^{1/2}$ characterizes the minimum distance between the point in $(x,y)$ plane and a coplanar point $(x_0,y_0)$ at the ellipse (refer to Fig. \ref{fig:ellipse_geometry} and the Appendix for a comprehensive elaboration on the determination of the trapping potential).   To investigate the behavior of a BEC in this quasi-1D elliptical waveguide, we set $\omega_z=\omega_\perp=29.34$ Hz, which corresponds to an oscillatory length of $l_   \perp=\sqrt{{\hbar}/(m\omega_\perp})=5\mu$m. 

In order to isolate and examine the influence of curvature on the condensate density distribution, maintaining a uniform cross-section along the waveguide is crucial. This requirement is fulfilled by the employed trapping potential described by Eq. (\ref{Eq:V_ext}), which establishes a parabolic trap $z$-direction and a waveguide in $(x,y)$ plane with a parabolic profile in perpendicular to the ellipse direction. The isolines of the potential form circles of constant radius along the waveguide, as shown in Figure \ref{fig:ellipse_geometry}.

In experimental practice, this potential can be easily realized by intersecting "sheet" and "elliptical" laser beams within an optical trap, following a technique akin to a well-established method for producing toroidal BECs \cite{Ramanathan2011}.  Vertical confinement is provided by a narrow red-detuned sheet beam, while the elliptical beam can be effectively created using optical traps that incorporate digital micromirror devices (DMDs), offering a feasible approach.




\begin{figure*}[ht]
\includegraphics[width=\textwidth]{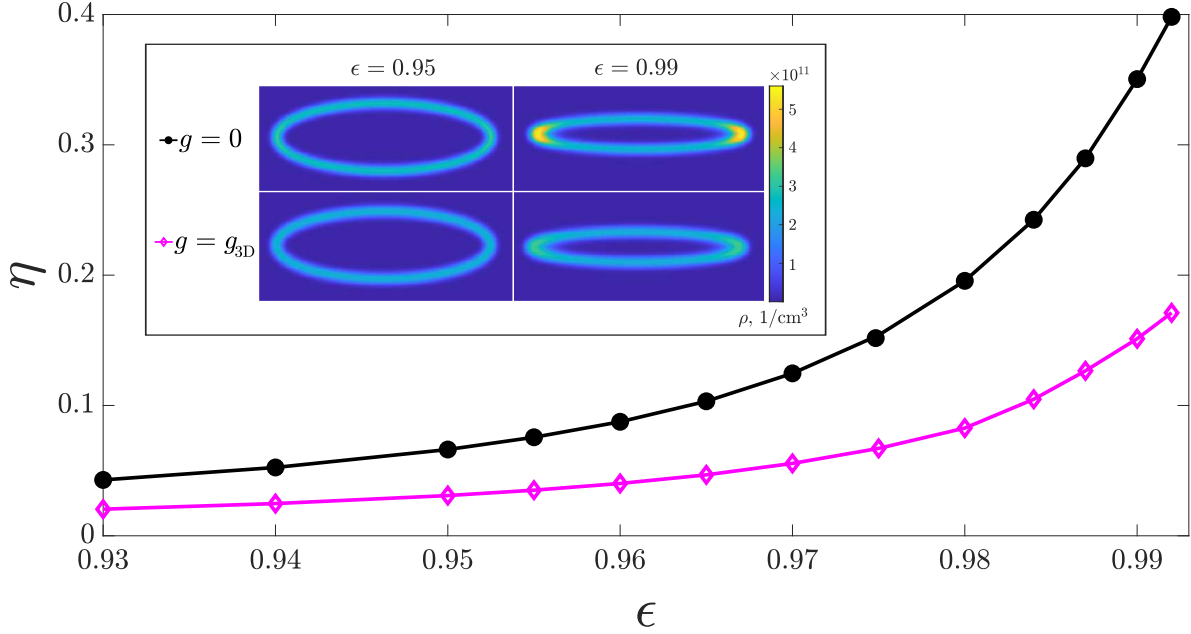}
\caption{Density modulation, $\eta$, as a function of eccentricity, $\epsilon$, for the ground state solution of the stationary 3D Gross-Pitaevskii equation. The modulation of density is depicted by the black curve with filled circles for the non-interacting case ($g=0$), while the magenta curve with diamonds represents the results for the repulsive interaction case ($g=g_{3D}$). Inset: Density distribution $|\Tilde{\Psi}|^2$ at the $z=0$ plane for the non-interacting case (upper row) and the case with repulsive interaction (lower row). Notably, density modulations arise due to curvature in regions with high curvature, and these modulations are reduced by strong repulsive interactions.} \label{fig:3D_eta_vs_epsilon}
\end{figure*}


We seek a steady-state solution of the form:
\begin{equation}
    \label{eq:stationary_psi}
    \Psi(\textbf{r},t)=\Tilde{\Psi}(\textbf{r})e^{-i\mu t/\hbar},
\end{equation}
where $\mu$ is the chemical potential. In general, the complex wavefunction $\Tilde{\Psi}=|\Tilde{\Psi}(\textbf{r})|e^{i\Phi(\textbf{r})}$ exhibits an inhomogeneous phase $\Phi(\textbf{r})$ with a circulation given by
\begin{equation}
    \label{eq:Circulation}
    \oint_C \nabla\Phi(\textbf{r})\cdot d\textbf{l} =2 \pi q,
\end{equation}
where the contour $C$ represents the ellipse defined by Eq. (\ref{eq:ellipseEq}), and $q$ is an integer denoting the topological charge of the wave function. For the ground state, $q=0$, while $q>0$ corresponds to a state with $q$ vortices, resulting in a counter-clockwise flow in the waveguide. The function $\Tilde{\Psi}(\textbf{r})$ satisfies the stationary GPE:
\begin{equation}
 \mu\Tilde{\Psi} = \left(-\frac{\hbar^2}{2m}
 \nabla ^{2}+V_{\text{ext}}(x,y,z)+g|\Tilde{\Psi} |^{2}
 \right) \Tilde{\Psi}   \label{eq:GPE_stationary}.
\end{equation}
To find the stationary states, we have employed the imaginary time propagation method, yielding numerical solutions. The inset in Fig. \ref{fig:3D_eta_vs_epsilon} illustrates typical examples of the density distributions for the ground states ($q=0$) in the non-interacting case ($g=0$) and the repulsive interaction case ($g=g_{3D}$).

To describe density modulation, we introduce a  parameter 
\begin{equation}
    \label{eq:Eta_def}
    \eta=\frac{n_a-n_b}{n_a+n_b},
\end{equation}
where $n_a$ and $n_b$ are the condensate densities $|\Tilde{\Psi}|^2$ calculated at the points ($x=a, y=0, z=0)$ and ($x=0, y=b, z=0)$ respectively. 
In Fig. \ref{fig:3D_eta_vs_epsilon}, we examine the relationship between density modulation, represented by $\eta$, and eccentricity for the ground state solution of the 3D GPE. The non-interacting case ($g=0$) is depicted by the black curve with filled circles, while the magenta curve with diamonds corresponds to the repulsive interaction case ($g=g_{3D}$). The inset provides a visualization of the density distribution at the $z=0$ plane for both scenarios.

The key finding from this analysis is the occurrence of density modulations induced by curvature and quantified by parameter $\eta$, particularly in regions characterized by high curvature. 
Remarkably, we observe that strong repulsive interactions lead to a reduction in density modulations, as evidenced by the comparison between the black and magenta curves. This indicates that the presence of repulsive interactions suppresses the amplitude of density  can be controlled by the strength of repulsive interactions.

\begin{figure*}[ht]
\includegraphics[width=\textwidth]{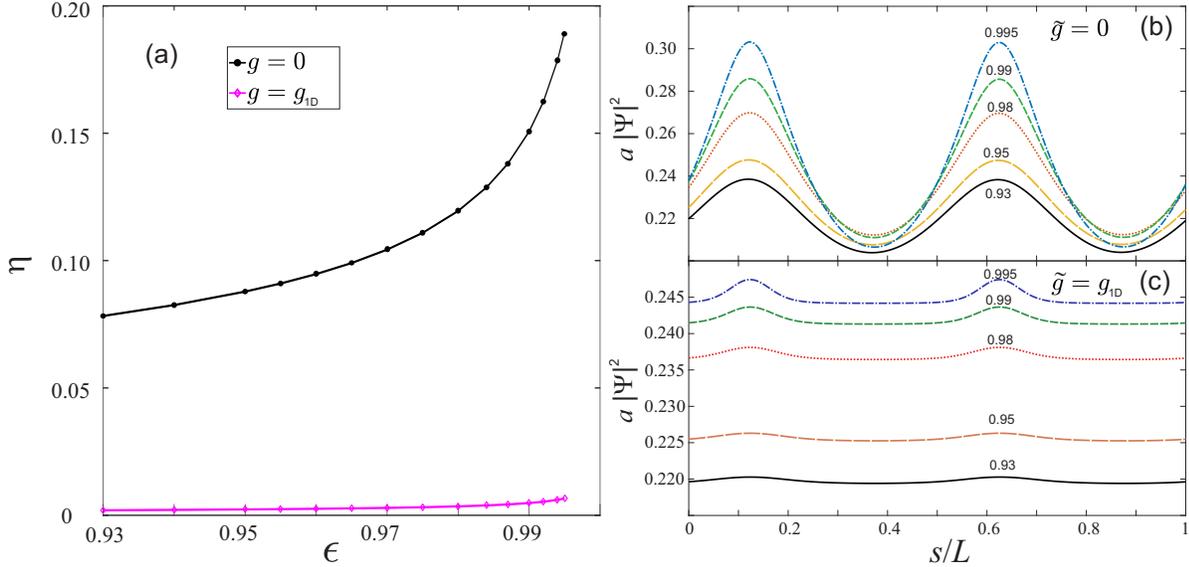}
\caption{(a) The modulation of the density, $\eta$, is shown as a function of eccentricity, $\epsilon$, for the 1D ground state solution of the non-polynomial nonlinear Schr\"odinger equation.
(b) The normalized condensate density, $a|\psi|^2$, is depicted for 1D stationary states without self-interaction ($\Tilde{g}=0$), as functions of the normalized coordinate $s/L$.
(c) The density profiles are displayed for the case with repulsive interaction ($\Tilde{g}=g_{1D}>0$).
In all cases, $L$ represents the perimeter of the ellipse, $a$ represents the length of the larger semi-axis of the ellipse, and the corresponding eccentricity values are indicated alongside the curves.
}
\label{fig:1D_eta_vs_epsilon}
\end{figure*}

\subsection{Effective 1D Model with Quantum Curvature Potential}
To compare the properties of the numerically obtained ground states, determined by solving the 3D stationary GPE, with an approximate effective linear potential induced by curvature, we use a non-polynomial  Schr\"odinger equation (NPSE)  \cite{SciPostPhysCore}. This model incorporates a quantum curvature-induced potential, which exhibits a double-well shape. The specific features of this potential are notably influenced by the eccentricity of the ellipse. Through this analysis, we can gain valuable insights into the impact of curvature on the properties of the ground state and the behavior of the condensate.

By assuming the factorization:
\begin{equation}
    \label{eq:factorization}
    \Psi(s,u,v, t) = \psi(s,t)\frac{e^{-\frac{u^2+v^2}{2\sigma(s,t)^2}}}{\sqrt{\pi}\sigma(s,t)},
\end{equation}
where $s$ represents the curvilinear abscissa (arc length), and $u$ and $v$ are the transverse plane coordinates, we obtain the NPSE as follows \cite{SciPostPhysCore}:
\begin{equation}\label{eq:dim_Psi}
i\hbar \partial_t \psi  = \left(-\frac{\hbar ^2}{2m}
\partial^2_s -\frac{\hbar^2 \kappa^2(s)}{8m}+\frac{\hbar^2}{2m}\frac{1}{\sigma^2}+\frac{m\omega_{\perp}^2}{2}\sigma^2 +\frac{\Tilde{g} (N-1)}{\sigma^2}|\psi|^2\right) \psi, 
\end{equation}
where $\sigma(s,t)$ is defined as
\begin{equation}
\sigma^2=l^2_{\perp}\sqrt{1+2a_s(N-1)|\psi|^2}.  
\label{eq:dim_sigma}
\end{equation}
In the above equations, we have the transverse length scale represented as $l_{\perp}=\sqrt{{\hbar}/{(m\omega_{\perp})}}=5 \mu m$, where $\omega_{\perp}$ denotes the transverse trapping frequency. The parameter $N=10^4$ indicates the number of particles in the condensate. In the 1D model, the interaction strength $\Tilde{g}$ for the repulsive interaction case is given by $\Tilde{g}=g_{1D}={2\hbar^2 a_s}/{m}$, where $a_s$ corresponds to the $s$-wave scattering length of the $^{87}$Rb atoms, and $m$ represents the atomic mass. On the other hand, for the non-interacting case, we have $\Tilde{g}=0$ (i.e., $a_s=0$). The normalization condition is defined as $\int_0^L|\psi|^2 ds=1$.

The curvature $\kappa$ of the ellipse, characterized by semi-axes $a$ and $b$, and eccentricity $\epsilon=\sqrt{1-b^2/a^2}$, can be expressed as a function of the polar angle $\varphi$:
\begin{equation}
\kappa(\varphi)=\frac{1}{a}\frac{\sqrt{1-\epsilon^2}}{\left(\sin^2{\varphi}+\sqrt{1-\epsilon^2}\cos^2{\varphi}\right)^{3/2}},  
\label{eq:dim_kappa}
\end{equation}
where we consider $a=100\mu m$ as the length of the ellipse's larger semi-axis. The arc length $s$ along the ellipse is given by:
\begin{equation}
s(\varphi)=a E(\varphi, \epsilon) = a \int_0^{\varphi}\sqrt{1-\epsilon^2\sin^2{\varphi'}}d\varphi',  
\label{eq:dim_s}
\end{equation}
with the perimeter of the ellipse defined as $L=aE(2\pi,\epsilon)$. Due to the symmetry, the curvature-induced potential $-\frac{\hbar^2 }{8m}\kappa^2(s)$ in Eq. (\ref{eq:dim_Psi}) implies effective double-well potential.


In Fig. \ref{fig:1D_eta_vs_epsilon}, we present the density and condensate profiles in an elliptic waveguide. Firstly, in panel (a), we plot the density modulation, $\eta$, as a function of eccentricity, $\epsilon$, for the 1D ground state solution of the NPSE.  Moving to panel (b), we depict the normalized condensate density, $a|\psi|^2$, for 1D stationary states without self-interaction ($g=0$), as functions of the normalized coordinate $s/L$, where $L$ represents the perimeter of the ellipse. This highlights the spatial variation of the condensate density along the waveguide. Additionally, in panel (c), we display the density profiles for the case with repulsive interaction ($g=g_{1D}>0$), showcasing the impact of interactions on the condensate distribution. Throughout the figure, the length of the larger semi-axis of the ellipse is denoted by $a$, and the corresponding eccentricity values are indicated alongside the curves. 

As shown in Figure \ref{fig:1D_eta_vs_epsilon}, the repulsive self-interaction effectively counteracts the curvature-induced density modulation. Notably, the 1D simulations exhibit qualitative agreement with the 3D GPE results presented in Figure \ref{fig:3D_eta_vs_epsilon}. However, it is important to consider the quantitative differences between the 1D and 3D models, particularly as the eccentricity increases. These differences become more pronounced due to the limitation of the factorization approximation (\ref{eq:factorization}), which is not able to describe the complex geometry of a real elliptical waveguide with a large eccentricity. While the quantum curvature potential successfully captures the main features of the density distribution, the 1D model with repulsive interaction noticeably underestimates the magnitude of the density modulation.

\section{Superfluid flows in curved elliptical waveguide}
\label{sec:s3}
Here we investigate the properties of the atomic flow in the 3D elliptic waveguide. Such states within a ring-shaped condensate can be induced using well-established methods using structured light \cite{Ramanathan2011}, stirring mechanisms \cite{2020PhRvA.102f3324E}, or phase imprinting procedures \cite{PhysRevA.97.043615}. First, we consider stationary solutions of the form (\ref{eq:stationary_psi}) for $q\ne 0$ and analyze the impact of the curvature on the phase distribution. Next, we address the question of the superflow stability and investigate the dynamics of the found stationary states.
\subsection{Stationary states with nonzero angular momentum}
\begin{figure*}[ht]
\includegraphics[width=6.8in]{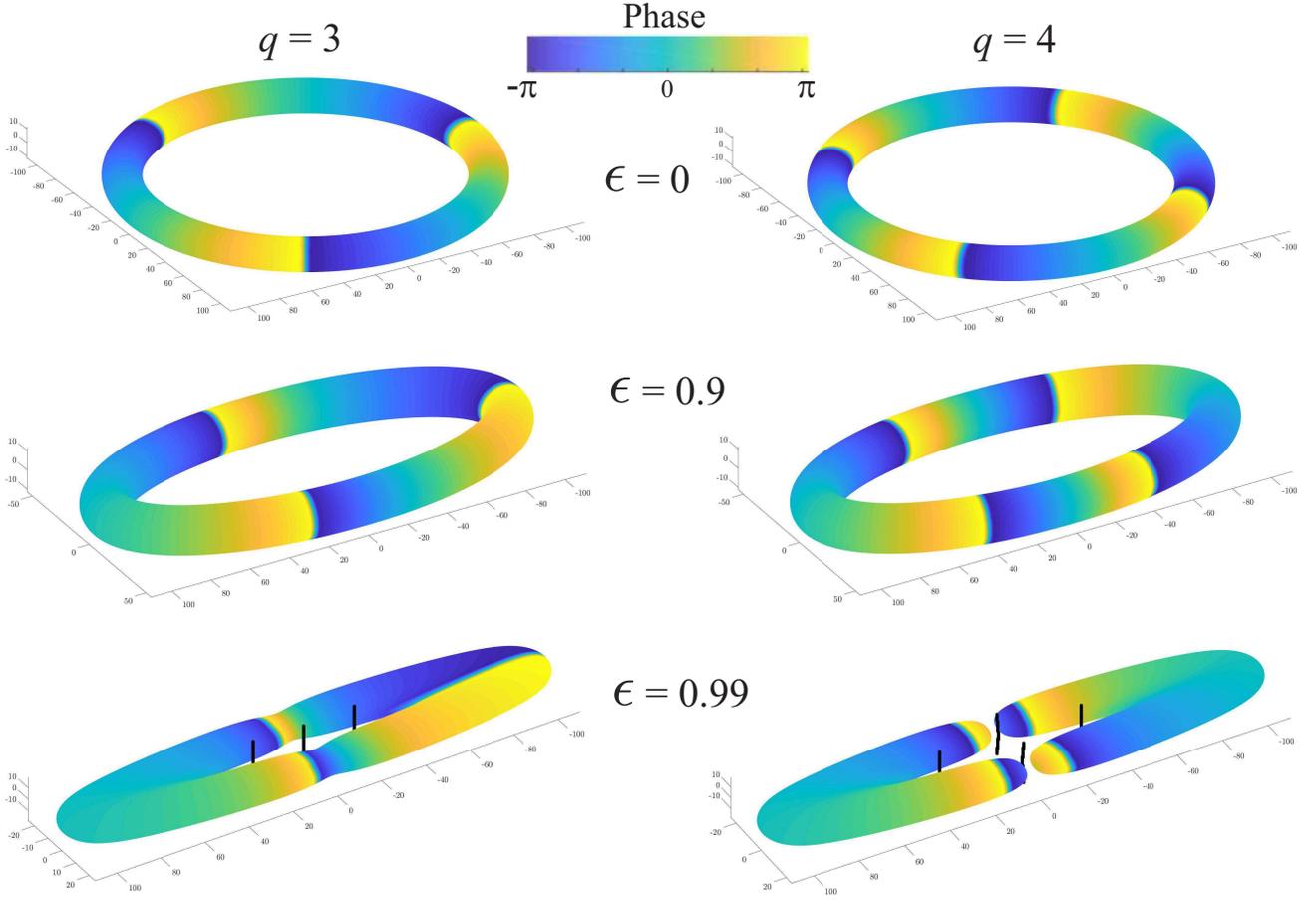}
\caption{\label{fig:density_isosurface}Density isosurface at 1\% of peak density for stationary states with topological charge $q=3$ (left) and $q=4$ (right) and varying values of the eccentricity $\epsilon$. The color scale represents the phase $\Phi$ of the wavefunction $\Tilde{\Psi}$ at the isosurfaces. Notably, the spatial distribution of the phase undergoes a transition from a homogeneous
phase gradient for $\epsilon = 0$ to a highly inhomogeneous phase distribution for waveguides with higher curvatures. Black lines in the lower row represent the vortex cores found numerically by the phase unwinding technique.}
\label{fig:s3s4}
\end{figure*}

We present numerical solutions of the form (\ref{eq:stationary_psi}) solving the stationary equation (\ref{eq:GPE_stationary}) for topological charges $q=1,2,3$ and 4. Figure \ref{fig:density_isosurface} illustrates representative examples of density isosurfaces, delineated at 1$\%$ of peak density, for stationary states with $q=3$ (left) and $q=4$ (right) for different eccentricity values. The color scale depicts the phase $\Phi(\textbf{r})$ of the wave function $\Tilde{\Psi}(\textbf{r})$ at the isosurfaces, providing insight into the associated topological charge $q$, which corresponds to the number of times the phase winds around $2\pi$ along an elliptic waveguide. Certainly, the total phase jump $2\pi q$ remains constant when eccentricity changes, but the spatial distribution of the phase  undergoes a transition from a homogeneous phase gradient for $\epsilon=0$ to a highly inhomogeneous phase distribution for waveguides with varying curvatures. The findings presented in Fig. \ref{fig:density_isosurface} highlight that an increased eccentricity $\epsilon$ leads to a comparatively uniform phase distribution within regions of higher curvature, effectively concentrating the phase variation in a compact region characterized by lower curvature. 

The vortex core positions were accurately determined using a numerical phase unwinding technique, and their locations are depicted by black lines in Fig. \ref{fig:density_isosurface}. However, it is important to note that the vortex core position is not shown for small eccentricity values due to limitations in our technique. Specifically, the determination of the vortex core position in an elliptic waveguide with a wide central hole, where the condensate is absent, is not feasible.

Our findings reveal an intriguing aspect concerning the remarkable phase variation observed in the vicinity of the vortex cores. Unlike the well-known solitonic vortices typically found in elongated condensates \cite{PhysRevA.65.043612}, our vortices reside not within the bulk of the condensate, but rather within the central hole of the toroidal waveguide. This distinction is highly significant. Moreover, the waveguide with toroidal topology enables the hosting of multiple vortices with a topological charge $q$. Consequently, the total $2\pi q$ phase jump along the waveguide becomes confined to a narrow region of lower curvature. This unique characteristic gives rise to a qualitatively distinct behavior compared to the previously studied dynamics of solitonic vortices in elongated, single-connected Bose-Einstein condensates.

Significantly, the use of an elliptic waveguide enables the accumulation of a substantial phase jump within a localized region, a remarkable achievement unattainable in a single-connected quasi-one-dimensional condensate. As a result, for $q \geq 3$, this accumulation of phase jump gives rise to the appearance of phase dislocations along the waveguide, reminiscent of domain walls and dark solitons observed in one-dimensional condensates (see Fig. \ref{fig:density_isosurface} for $\epsilon=0.99$). Notably, such strong phase variation is accompanied by essential density redistribution.

In the subsequent section, we analyze the dynamics of the superflows in the curved waveguide, shedding light on the connection between the phase jump and density behavior. Our findings demonstrate that the phase jump can lead to the formation of regions with density nodes, akin to domain walls between regions exhibiting a phase difference of $\pi$.

\subsection{Dynamics and stability of the superflows in a curved elliptic waveguide.}
In the study of nonequilibrium phenomena, such as the nucleation of vortices and decay of the superflow, the role of dissipative effects cannot be overstated as they play a critical role in the relaxation process towards an equilibrium state. Dissipation provides the mechanism by which vortex lines either drift towards the outer edge of the condensate, where vortices eventually decay, or become pinned in the central hole of a ring-shaped condensate. The relaxation of the vortex core position towards the local energy minimum gives rise to the formation of a metastable superflow.

Dissipative effects manifest themselves in a trapped condensate through interactions with a thermal cloud and can be phenomenologically captured by the dissipative GPE. This equation describes the dynamics of the macroscopic wave function for a system of weakly interacting degenerate atoms in proximity to thermodynamic equilibrium, subject to weak dissipation \cite{Pitaevskii1958,Choi98}:
\begin{equation}
 (i-\gamma)\hbar\frac{\partial \Psi }{\partial t}= \left(-\frac{\hbar^2}{2m}
 \nabla ^{2}+V_{\text{ext}}(x,y,z)+g_{3D}|\Psi |^{2}
 -\mu\right) \Psi \label{eq:dampedGPE3D}.
\end{equation}
Here $\gamma\ll 1$ is a phenomenological dissipation parameter and $\mu$ is the chemical potential of the state with $N$ atoms.

The $\gamma$ parameter plays a crucial role in determining the relaxation time of the vortices within the system. Specifically, a larger value of $\gamma$ corresponds to a shorter timescale for vortices to migrate from the high-density region of the condensate annulus to the low-density periphery. In the subsequent analysis, we make the assumption of a constant dissipative parameter $\gamma$ and set its value to $\gamma=0.03$, disregarding any potential position dependence. Importantly, we have verified that our key findings remain qualitatively unchanged irrespective of the specific value chosen for $\gamma$, as long as $\gamma \ll 1$. 

To emphasize the significant impact of curvature on phase accumulation, we focus on the dynamics of an elliptic BEC with a high eccentricity, specifically with an eccentricity value of $\epsilon=0.99$. Through extensive numerical simulations, we have observed complex evolution in the system, which we describe below. It's important to note that, even in the presence of dissipation, the total topological charge is conserved throughout the entire space, including the bulk of the BEC and the surrounding region.
While the dissipative process causes energy reduction, leading to the transformation into lower-energy states with lower angular momentum, it's crucial to emphasize that vortex cores do not disappear or suddenly emerge. Instead, they can drift to the edge of the condensate or annihilate with their oppositely charged counterparts.

Our numerical simulations indicate that not only the ground state ($q=0$) but also the single-charged ($q=1$) and double-charged ($q=2$) superflows remain stable over long time scales, even for high eccentricity ($\epsilon=0.99$). These superflows maintain their coherent flow patterns without significant changes.

However, when considering higher-charged superflows with $q\geq 3$, we observe a complex series of dynamic transformations between different states. These higher-charged superflows experience intricate changes in their flow patterns and topological structures as time progresses. The evolution of these superflows involves transitions between various states, leading to rich and intricate dynamics.

 \begin{figure*}[ht]
\includegraphics[width=\textwidth]{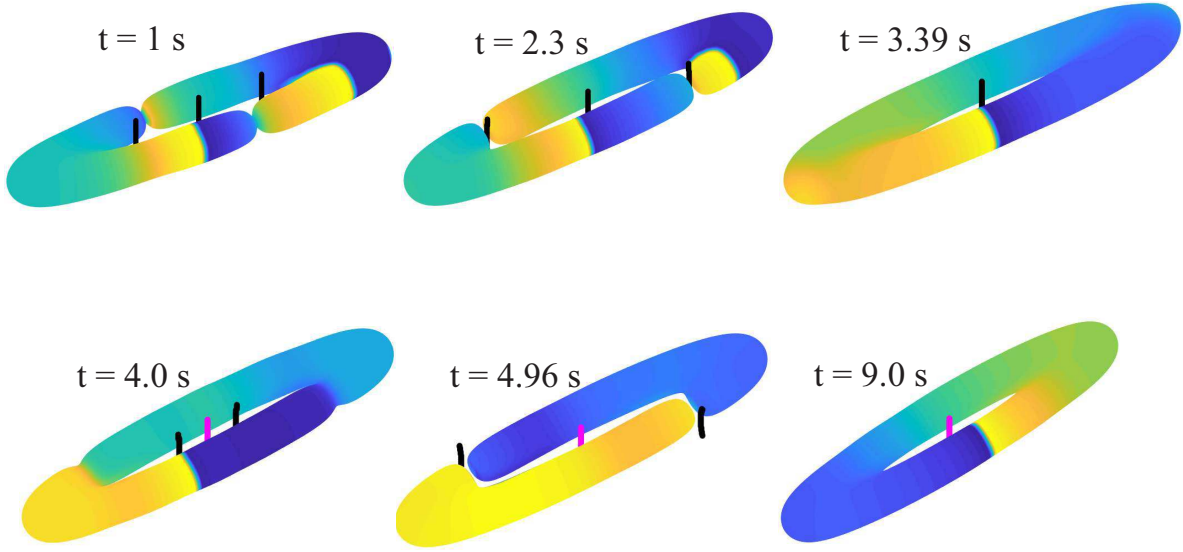}
\caption{Snapshots of density and phase during the dissipative evolution of the superflow with $\gamma=0.03$. The density isosurface and phase of the wavefunction are shown with the same scales as in Fig. \ref{fig:density_isosurface}. The cores of vortices (antivortices) are depicted by black (magenta) lines.  
The initial state, represented by a topological charge of $q=+3$ and an eccentricity of $\epsilon=0.99$, is displayed in Fig. \ref{fig:s3s4}. Notably, two weak links  emerge and propagate in an anti-clockwise direction, aligning with the superflow (as observed in the snapshot at $t=1$ s). Two vortex lines pass through the weak links and subsequently escape (as shown at $t=2.3$ s). Meanwhile, the central vortex line remains at the center of the ring, leading to the formation of a superflow with a topological charge of $q=+1$, as observed at $t=3.39$ s.
During further evolution, the central vortex with $q=+1$ transforms into a stationary central antivortex with $q=-1$, while the two $q=+1$ vortices move aside the center towards the weak links and escape. Notably, the final state exhibits a stable clockwise superflow, characterized by the phase and topological charge of an {antivortex} ($q=-1$). Remarkably, the direction of the superflow in this final state is reversed compared to the initial state, which had an anti-clockwise flow direction.
}\label{fig:s3_evolution}
\end{figure*}
 \begin{figure*}[ht]
\includegraphics[width=\textwidth]{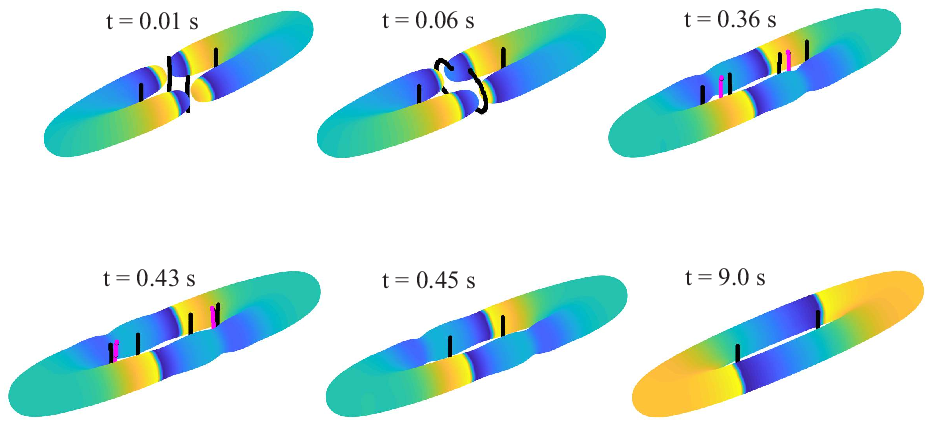}
\caption{The same as in Fig.  \ref{fig:s3_evolution} for $q=+4$. The initial state is shown in Fig. \ref{fig:s3s4}. At $t=0.06$ s, a phase slip occurs, transitioning the state to $q=2$, accompanied by the escape of two vortices through the weak links. Subsequently, at $t=0.36$ s, in addition to the two vortex lines, a pair of dipoles emerges. Notably, at $t=0.43$ s, a 'recharging' process takes place, where the two antivortices decouple from the dipoles and annihilate with a pair of vortex lines. The long-term evolution demonstrates the establishment of a stable $q=2$ flow.}
\label{fig:s4_evolution}
\end{figure*}

 Figure \ref{fig:s3_evolution} illustrates the dissipative evolution of the superflow with the initial state exhibiting a topological charge of $q=3$. Notably, the snapshots demonstrate the emergence of two weak links (localized regions of reduced superfluid density in the
condensate annulus)  that propagate in an anti-clockwise direction, aligning with the flow (as observed at $t=1$ s). 
At $t=2.3$ s, two vortex lines traverse through the weak links and subsequently escape, as illustrated in Fig. \ref{fig:s3_evolution}. The central vortex line remains at the center of the ring, leading to the formation of a superflow with a topological charge of $q=+1$. This state can be observed at $t=3.39$ s. 
Then the single-charged central vortex undergoes splitting at $t=4.0$ s. The splitting results in the formation of two vortex lines (depicted by black lines) and one antivortex (represented by the magenta line in Fig. \ref{fig:s3_evolution}). Subsequently, during further evolution, an antivortex with a topological charge of $q=-1$ remains close to the center, while the two vortex lines with a topological charge of $q=+1$ move towards the weak links and escape. This dynamic is clearly captured in the snapshot for $t=4.96$ s. 
A noteworthy transient state is observed at $t=4.96$ s in Fig. \ref{fig:s3_evolution}, where the condensate splits into two fragments with nearly constant phases close to zero and $\pi$, respectively. These fragments are separated by a pair of domain walls that exhibit condensate density nodes. At first glance, this state might appear to have zero vorticity, since the condensate phase resembles the phase structure of a pair of dark solitons rather than a vortex phase. However, a detailed analysis of the phase reveals an additional sharp phase gradient in the internal region of the trap. As a result, there is phase circulation along the closed path surrounding the center of the ellipse, which leads to a phase winding of $-2\pi$. This phase winding corresponds to the presence of an antivortex line, depicted by the magenta color near the center of the trap.
 
Remarkably, the final state of the system corresponds to a stable clockwise flow characterized by the phase and topological charge of an {antivortex} with $q=-1$. Consequently, the direction of the superflow in this final state is opposite to the anti-clockwise direction of the initial state's flow.
  

In Fig. \ref{fig:s4_evolution}, we present snapshots that illustrate the dissipative evolution of the initial state characterized by a topological charge of $q=4$ (see Fig. \ref{fig:s3s4}  for an eccentricity value of $\epsilon=0.99$). At $t=0.06$ s, a phase slip occurs, resulting in a transition of the state to a topological charge of $q=2$, accompanied by the escape of two vortices through the weak links in the density distribution. It is noteworthy that the escaping vortex lines exhibit significant curvature, providing visual evidence of the three-dimensional phase structure of the phase slip process. As time progresses, at $t=0.36$ s, the system exhibits the emergence of a pair of dipoles in addition to the two vortex lines. Remarkably, at $t=0.43$ s, a 'recharging' process takes place, where the two antivortices decouple from the dipoles and annihilate with a pair of vortex lines, leading to a rearrangement of the superflow configuration. The long-term evolution of the system demonstrates the establishment of a stable superflow with a topological charge $q=2$.

The supplementary materials feature animations showcasing the dynamics of two specific examples depicted in Figs. \ref{fig:s3_evolution} and \ref{fig:s4_evolution}. These animations offer a comprehensive visualization of the evolution of the superflow, providing detailed insights into the  evolution of the condensate density and phase. 

Therefore, our investigations of the dynamics of elliptic BEC with high eccentricity ($\epsilon=0.99$) have revealed that single-charged and double-charged flow maintain stability, while higher-charged superflows $q\ge 3$ undergo intricate transformations between different states. These findings illustrate the profound influence of curvature on the behavior of superflows in the system and highlight the complex nature of their dynamics.

\section{Conclusions}\label{sec:Conclusions}
We have conducted a comprehensive investigation of the influence of curvature on the phase and density of atomic Bose-Einstein condensates confined in elliptical trapping potentials. Our study has yielded several key findings.

Firstly, we have analyzed the quasi-1D elliptical trap with varying eccentricities and observed that the curvature of the waveguide has a significant impact on the density distribution of the stationary ground states. Utilizing an approximate 1D model based on the non-polynomial nonlinear Schr\"odinger equation  with an effective curvature-induced potential, our results exhibit a qualitative agreement with 3D numerical simulations. Specifically, we have observed the emergence of local density peaks in the region with the highest curvature, which is induced by the ellipticity of the waveguide. However, the presence of repulsive self-interaction counteracts the curvature-induced density modulation.

Additionally, our investigation into the phase behavior of elliptical waveguides with superflows has revealed that waveguides with high eccentricity provide phase gradient accumulation in regions with lower curvature. To explore the evolution of superflows with different topological charges, we have conducted a series of numerical simulations using the damped 3D GPE. Our findings demonstrate the stability of ground states and superflows with $q=1$ and $q=2$, even for significant eccentricities. However, for initial states with a topological charge $q\ge 3$, we have observed the further development of an inhomogeneous phase distribution along the waveguide, characterized by emergence of vortex-antivortex pairs, phase jumps and density nodes associated with dark solitons.  
Specifically, the initial state with $q=3$ decays into the $q=1$ state, which subsequently transforms into a pair of moving $q=1$ vortices and a central antivortex with $q=-1$, before ultimately transitioning into the antivortex state with $q=-1$ so that the final state exhibits a counter-propagating flow direction compared to the initial state.

We believe that our findings provide important insights into the behavior of atomic Bose-Einstein condensates confined in curved waveguides and their potential applications. The ability to control and manipulate the properties of these systems holds great promise not only for advancing our understanding of fundamental physics but also for the development of innovative quantum technologies based on coherent matter waves. 

Since the atomtronic circuits inherently involve curved waveguides, the ability to govern the distribution of condensate density and phase holds crucial importance in the development of quantum sensors and information processing systems based on atomic BECs. The precise control over these parameters within curved waveguides could facilitate the design of advanced quantum devices with enhanced sensitivity and functionality.

Extending the scope of our current research, it would be relevant to explore the interplay between curvature and interaction in quasi-2D structures. This extension could offer valuable insights into the remarkable behaviors of atomic condensates under varying dimensional constraints.  Additionally, a promising avenue for further analysis involves the investigation of vortex states within quasi-2D systems characterized by substantial curvature. Experimental validation of these states could be pursued through existing and ongoing experiments utilizing quantum bubbles in microgravity environments.


\section{Acknowledgment}
The authors thank Gerhard Birkl for useful discussions. Y.N. acknowledges support by the Austrian Science Fund (FWF) [Grant No. I6276]. L.S. is partially supported by the European Quantum Flagship Project "PASQuanS 2" and by the European Union-NextGenerationEU within the National Center for HPC, Big Data and Quantum Computing [Project No. CN00000013, CN1 Spoke 10: “Quantum Computing”]. A.Y. and Y.N. acknowledge support from  the National Research Foundation of
Ukraine through Grant No. 2020.02/0032. L.S. and A.Y. acknowledge support from the BIRD Project "Ultracold atoms in curved geometries" of the University of Padova and from “Iniziativa Specifica Quantum” of INFN.

\section{Appendix}

\subsection{Trapping potential}
To determine the coordinates $(x_0,y_0)$ of the corresponding point on the ellipse, we employ the method of Lagrangian multipliers, minimizing the distance from a given point $(x,y)$ to the ellipse under the constraint given by the ellipse equation:
\begin{equation}\label{eq:ellipseEq}
\frac{x_0^2}{a^2}+\frac{y_0^2}{b^2}=1.    
\end{equation}
This leads to the following expressions for $(x_0,y_0)$:
\begin{equation}\label{eq:x0y0}
x_0 = \frac{xa^2}{a^2+\lambda}, y_0 = \frac{yb^2}{b^2+\lambda},    
\end{equation}
where $\lambda$ is the solution of the equation
\begin{equation}\label{eq:lambdaEq}
    \frac{x^2a^2}{(a^2+\lambda)^2}+\frac{y^2b^2}{(b^2+\lambda)^2} = 1.
\end{equation}

Thus, we obtain $R(x,y)=\left[(x-x_0)^2+(y-y_0)^2\right]^{1/2}$, with $(x_0,y_0)$ determined using Eq. (\ref{eq:x0y0}). The parameter $\lambda$ is obtained by solving Eq. (\ref{eq:lambdaEq}). Since the corresponding quartic equation for $\lambda$ cannot be solved analytically in general case, we find the roots of Eq. (\ref{eq:x0y0}) numerically.


\subsection{Numerical methods}
We employed the Split-Step Fourier Method (SSFM) to solve the dynamical Gross-Pitaevskii equation (GPE) \cite{agrawal2019nonlinear}. This approach splits the operators in the GPE into linear and nonlinear components.
The linear operator evaluation is performed in the frequency domain, efficiently utilizing the Fast Fourier Transform algorithm (FFT).
Furthermore, the SSFM is capable of providing solutions for stationary equations. By substituting $\Delta t \rightarrow -i \Delta t$ and normalizing each function, a converging sequence toward the system's eigenstate is attained (Imaginary Time Propagation technique). This method can be readily extended for  numerical solutions of dissipative GPE. Since the 3D calculating schemes are numerically very demanding, they have been implemented for graphics processing units (GPUs) using    CUDA, enabling a very high degree of parallelization. 

We employed the following approach to determine the coordinates of vortex cores, as well as their rotation direction and topological charge. For each grid point, we consider three perpendicular planes.
In these planes, we analyze eight grid nodes surrounding the considered point, excluding those forming the boundary. We initialize the phase value of the wave function at the starting point as zero. Since phase values naturally span the interval $(-\pi, \pi)$, a phase jump is observed when this interval's boundary is crossed. The topological charge can be inferred from the number and sign of these phase differences.

It's worth noting that a similar methodology was previously employed for phase unwrapping in the context of a 2D problem, as described in Ref. \cite{caradoc-davies2000vortex}.

\section{Supplemental Materials}
 In the Supplemental Materials, we present animations illustrating the phase and density evolution of trapped Bose-Einstein condensates (BECs) for two examples featuring distinct initial topological charges. A comprehensive analysis of each example can be found in the main text (see Fig. 4 in the main text for $q=3$ and Fig. 5 for $q=4$). 
 
\begin{figure*}[ht]
\includegraphics[width=\textwidth]{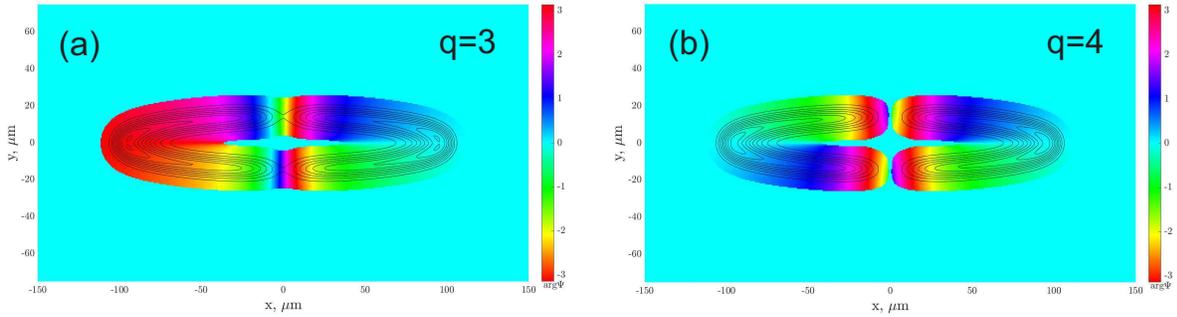}
\caption{Color-coded phase of the condensate in $z=0$ plane combined with
density isolines for the stationary states with superflow. (a) Topological charge $q=3$. (b) Topological charge $q=4$. }
\label{fig:SM1}
\end{figure*}

 In addition, in Fig. \ref{fig:SM1}  we illustrate the initial state for both cases in the form of the color-coded phase combined with isolines of the density distribution in $z=0$ plane.

Simulations of damped 3D GPE are performed within an elliptical waveguide characterized by an eccentricity of $\epsilon=0.99$ and a dimensionless damping parameter of $\gamma=0.03$. The upper subplot of the animations demonstrates the temporal evolution of the density distribution in the $z=0$ plane, while the lower subplot depicts the corresponding variation in the wave-function phase. The green dots represent the positions of vortex cores,  with positive circulation ($q=1$), while the magenta dots correspond to antivortex cores, signifying negative circulation ($q=-1$).

\subsection{Dynamics of the superflows with $q=3$}
The first video (file SMq3.avi) corresponds to the case of $q=3$ initial topological charge.  Initially, three vortex line perpendicular to the $(x,y)$ plane are imprinted and obtained ground state has noticeable density dips (weak links). The phase gradient is bigger in the center of the ellipse, while on the peripheries, the phase is almost uniform. Starting from this initial condition, the system shows the repulsion of the similarly charged vortex lines, and the weak links move counterclockwise. The two antivorteces with cores marked by magenta dots approach several times, enter inside the ellipse through the bulk gates and annihilate with two vortices ($t=1.61$ s). Vortex-antivortex pair is recombined again, but leave through the weak links ($t=2.55$ s), which heal in a while ($t=3.42$ s). The central vortex line remains in the system, leading to the formation of the flow with $q=1$. The weak links appear again, moving this time in the clockwise direction. The phase distribution significantly transforms, showing two regions with almost uniform distributions differ by $\pi$. 

The single charge vortex splits to a triple of two vortices and one antivortex which is this time located in the center ($t=3.89$ s). Another vortex-antivortex pair appears ($t=4.06$ s), but then annihilates ($t=4.12$ s). Two vortices approach the central antivortex and annihilate, leading to one vortex remain in the center ($t=4.27$ s). Two vortex-antivortex pairs appear ($t=4.33$ s). Antivorteces annihilate, with the central vortex ($t=4.71$ s) remaining antivortex in the center. Two vortices escape through the weak links ($t=4.97$ s). Then again the central antivortex forms a triple with vortex and two antivorteces which start moving towards the weak links. 
The phase again forms two fragments separated by a pair of domain walls that exhibit condensate density
nodes. At first glance, this state might appear to have zero vorticity, since the condensate phase resembles the phase structure of a pair of dark solitons rather than a vortex phase. However, a detailed analysis of the phase reveals an additional sharp phase gradient in the internal region of the trap. As a result, there is phase circulation along the closed path surrounding the centre of the ellipse, which leads to a phase winding of $-2\pi$. This phase winding corresponds to the presence of an antivortex line near the centre of the trap. After a rather long-term relaxation process system reaches  the equilibrium state with $q=-1$. Notably, the direction of the superflow in this final state is opposite to the anti-clockwise direction of the initial state's flow.

\subsection{Dynamics of the superflows with $q=4$}
In the second video (file SMq4.avi), the system evolves under the same parameter set as the first case, with the exception of the initial topological charge $q=4$. Notably, the dynamics demonstrate the rapid escape of two vortices through the weak links, occurring at $t=0.07$ s. Subsequently, at $t=0.36$ s, a pair of dipoles emerges in addition to the two vortex lines. However, the antivortices of the dipoles quickly annihilate, resulting in a stable pair of two vortices at $t=0.44$ s.

\section*{References}
\bibliography{ref}    
\end{document}